\documentclass[final,5p,twocolumn,12pt]{elsarticle}

\usepackage{graphicx}
\usepackage{amssymb}
\usepackage{amsmath}

\biboptions{sort&compress}

\usepackage{listings}
\usepackage{fancybox}
\newcommand{\code}[1]{\texttt{#1}}
\PassOptionsToPackage{normalem}{ulem}
\usepackage{ulem}
\usepackage{subfig}
\usepackage{multirow}
\usepackage{colortbl}
\definecolor{lightgray}{gray}{0.8}

\newcounter{bla}

\journal{Computer Physics Communications}

\begin{document}

\newcommand{\ME}{\begin{tt}Maxent\end{tt}}
\newcommand{\MEsp}{\begin{tt}Maxent\end{tt} }
\begin{frontmatter}

\title{Implementation of the Maximum Entropy Method for Analytic Continuation}

\author[a]{Ryan Levy}
\author[a]{J.P.F. LeBlanc}
\author[a]{Emanuel Gull\corref{author}}

\cortext[author] {Corresponding author.\\\textit{E-mail address:} egull@umich.edu}
\address[a]{Department of Physics, University of Michigan, Ann Arbor, MI 48109, USA}

\begin{abstract}
We present {\ME}, a tool for performing analytic continuation of spectral functions using the maximum entropy method. The code operates on discrete imaginary axis datasets (values with uncertainties) and transforms this input to the real axis. The code works for imaginary time and Matsubara frequency data and implements the `Legendre' representation of finite temperature Green's functions. It implements a variety of kernels, default models, and grids for continuing bosonic, fermionic, anomalous, and other data. Our implementation is licensed under GPLv2 and extensively documented. This paper shows the use of the programs in detail.
\end{abstract}


\end{frontmatter}

{\bf PROGRAM SUMMARY}

\begin{small}
\noindent
{\em Manuscript Title: }   Implementation of the Maximum Entropy Method for Analytic Continuation                                    \\
{\em Authors: }                       Ryan Levy, J.P.F. LeBlanc, Emanuel Gull                         \\
{\em Program Title: }      maxent                                    \\
{\em Journal Reference:}                                      \\
{\em Catalogue identifier:}                                   \\
{\em Licensing provisions:}     GPLv2                               \\
{\em Programming language: C++}                                   \\
{\em Operating system:} Tested on Linux and Mac OS X                                    \\
{\em RAM:} 10 MB -- 200 MB                                              
{\em Classification:} 4.9 \\
{\em External routines/libraries:} ALPSCore \citep{ALPS20}\cite{ALPSCore}, GSL, HDF5 \\
{\em Nature of problem:
}The analytic continuation of imaginary axis correlation functions to real frequency/time variables is an ill-posed problem which has an infinite number of solutions. \\
   \\
{\em Solution method: }The maximum entropy method obtains a possible solution that maximizes entropy, enforces sum rules, and otherwise produces `smooth' curves.  Our implementation allows for input in Matsubara frequencies, imaginary time, or a Legendre expansion. It implements a range of bosonic, fermionic and generalized kernels for normal and anomalous Green's functions, self-energies, and two-particle response functions.\\
   \\
{\em Running time: }10s - 2h per solution\\
   \\

\end{small}

\section{Introduction}
Analytic continuation of numerical data is a standard problem in condensed matter physics.
It primarily appears when correlation functions of a many-body problem, computed in an imaginary time statistical mechanics formulation, need to be interpreted as response or spectral functions on the real axis.
While imaginary time (or the Fourier transform, Matsubara frequency) correlation functions are naturally obtained in numerical simulations such as quantum Monte Carlo lattice \cite{BSS81} and impurity solvers  \cite{Rubtsov05,Werner06,Gull08_ctaux,Gull11_RMP},  their real axis counterparts that correspond to response functions, which are measured in experiment, are not typically accessible to numerical techniques. 

At the heart of this is that the continuation from the imaginary axis to the real axis is exceptionally ill conditioned, such that small fluctuations of the input data (either from statistical Monte Carlo noise or a truncation of the accuracy to finite precision numbers) lead to large fluctuations of the output data, rendering any direct transformations useless in practice.
Several alternatives have been proposed, among them the construction of rational polynomial functions (Pad\'{e} approximants) \citep{beach:2000,padeBook,Zhu2002}, a constrained optimization procedure \cite{nikolay:jetp}, a stochastic analytic continuation method \cite{sandvik:1998}, and a stochastic analytic inference method \cite{fuchsSAI}. The standard method, however, is the so-called maximum entropy method (MEM) \cite{jarrell1996bayesian,bryanMaxent}, for which we provide an implementation in this paper. Our implementation, \ME, is part of the ALPS applications \cite{ALPS,ALPS20,ALPS_DMFT} and makes use of the core ALPS libraries \cite{ALPSCore}. In the following, we will briefly review the formalism (referring the reader to Ref.~\cite{jarrell1996bayesian} and the original literature for more details), introduce our implementation, and illustrate its usage with examples.

\section{Analytic Continuation}
\subsection{Analytic Continuation Formalism}
We start our considerations with the imaginary time Green's function $G(\tau) = -\langle c(\tau) c^\dagger (0)\rangle$, which is a continuous function for $0<\tau<\beta$, and is periodic for bosonic and anti-periodic for fermionic systems within $\tau\in [0,\beta]$. In this notation, $c$ denotes an annihilation operator, $c^\dagger$ a creation operator, and the time-dependence of the operator and its expectation value are to be interpreted in the usual sense \cite{mahan2013many}.
Imaginary time Green's functions of this type are the fundamental objects that most QMC methods produce as a simulation output. 
The Green's function in $\tau$ can be related to the Green's function on the imaginary frequency axis through a Fourier transform
\begin{equation}
G(i\omega_n)=\int_0^\beta e^{i\omega_n\tau}G(\tau).
\end{equation}
The `Matsubara' frequencies $i\omega_n$ come from poles of the distribution functions and are defined as $i\omega_n = 2\pi(n+\frac{1}{2})/\beta$ for fermionic and $i\omega_n=2\pi n/\beta$ for bosonic operators.

For the rest of the article, we assume that these Green's functions are not known to arbitrary precision. Rather, we work with a truncation of the Green's function to $N$ components, which are obtained by averaging a set of $M$ estimates for each component, $G^{(i)}_n$, that are independent and Gaussian distributed so that if there are $M$ samples for each $n$, the estimate for the Green's function is given by 
\begin{equation}
G_n = \dfrac{1}{M} \sum_{j=1}^M G_n^{(j)}.
\end{equation}
Different components $n$ and $m$ of the Green's function may be correlated. This is encapsulated in the covariance matrix $C_{nm}$, which is estimated as 
\begin{equation}
C_{nm} = \dfrac{1}{M(M-1)}\sum_{j=1}^M(G_n-G^{(j)}_n)(G_m-G^{(j)}_m).
\label{eq:cov_matrix}
\end{equation}

In the case of fermions, the Matsubara frequency Green's function $G(i\omega_n)$ and its imaginary time counterpart $G(\tau)$ are related to a real frequency Green's function $G(\omega)$ via
\begin{align}
G(i\omega_n) &= \frac{-1}{\pi}\int_{-\infty}^{\infty}  \frac{d\omega \text{Im}\left[G(\omega)\right]}{i\omega_n - \omega},\label{omegacont}\\
G(\tau_n) &= \frac{1}{\pi}\int_{-\infty}^{\infty}  \frac{d\omega \text{Im}\left[G(\omega)\right]e^{-\tau_n\omega}}{1+e^{-\beta\omega}}, \label{taucont}
\end{align}
where $\tau$ has been discretized in some manner to $N$ points.
The imaginary part of the Green's function that appears in the numerators of Eqs.~\eqref{omegacont} and \eqref{taucont} defines the spectral function
\begin{equation}
A(\omega)=-\dfrac{1}{\pi}\mbox{Im}\left[G(\omega)\right].
\end{equation}
Obtaining $G(\omega)$ and $A(\omega)$ in addition to related quantities for bosonic and other response functions, as well as self-energies, is the main purpose of this paper.

We can formulate Eq.~\ref{taucont} as 
\begin{align}
G_n=G(\tau_n)&=\int_{-\infty}^{\infty} d\omega\thinspace A(\omega)K_n(\omega),
\label{eq:G=KA}\\
K_n(\omega) &= K(\tau_n,\omega)=-\dfrac{e^{-\tau_n\omega}}{1+e^{-\omega\beta}},
\end{align}
where $K_n$ is the `kernel' of the analytic continuation, here for a transformation of a fermionic Green's function from imaginary time to real frequencies. Kernels for other distribution functions and imaginary axis representations are listed in Sec.~\ref{sec:kernels},  Tables~\ref{tab:frequency-kernels} and \ref{tab:time-kernels}.  

Given a candidate spectral function $A(\omega)$ on the real axis and the associated kernel, the imaginary axis Green's function can be evaluated using Eq.~\eqref{eq:G=KA} to create an estimate $\bar{G}_n$, a process known as  a back-continuation. To calculate the consistency of a spectral function $A(\omega)$ with the imaginary axis data $G_n$, one can define a ``goodness of fit" quantity $\chi^2$
\begin{equation}
\chi^{2}=\sum_{n,m}^{M}(\bar{G}_{n}-G_n)^{*}C_{nm}^{-1}(\bar{G}_{m}-G_{m}),
\label{eq:chi2}
\end{equation}
where $C_{nm}$ is defined in Eq.~\eqref{eq:cov_matrix}. Consistency of $A$ with $G_n$ within errors given by $C_{nm}$ is achieved for $\chi^2~\sim~M$.
If the input data is uncorrelated then only the diagonal elements of the covariance matrix are non-zero, in which case $\chi^2$ takes the form
\begin{equation}
\chi^{2}=\sum_{n}^{M}\frac{(\bar{G}_{n}-G_n)^2}{\sigma_n^2}
\end{equation}
where $\sigma_n$ is the standard error in $G_n$.

\subsection{Inversion of the Kernel}
To computationally solve for $A(\omega)$ in Eq.~\eqref{eq:G=KA}, the simplest method to employ is a least squares fitting routine, which attempts to minimize a functional $Q=\chi^2$ with $\chi^2$ described in Eq.~\eqref{eq:chi2}. While back-continuation is a straightforward procedure that gives a unique result, the inversion of  Eq.~\eqref{eq:G=KA} is ill conditioned, {\it i.e.} there are many solutions $A$ that satisfy $G=KA$ to within the uncertainty given by $C_{nm}$.

\subsection{Maximum Entropy Method}
Instead of least-squares fitting, analytic continuation algorithms impose additional criteria on $A$, such as smoothness \cite{nikolay:jetp}, in order to reduce the space of acceptable solutions. In the maximum entropy method an ``entropy'' term, $S$, is also considered to help regularize the solutions, such that the functional to minimize becomes
\begin{equation}\label{eq:Q}
Q=\dfrac{1}{2}\chi^{2}-\alpha S[A],
\end{equation}
where the factor of $1/2$ is added for mathematical convenience and we have introduced a Lagrange multiplier $\alpha>0$ here, in order to control the competition between $\chi^2$ and $S$. 

Shannon entropy \cite{shannonentropy} is used in order to minimize spurious correlations between data \citep{gullSkilling}; this restricts its application to  cases where the resulting spectral function is finite and positive (or can be transformed to be), so it can be treated as a probability density \cite{jarrell1996bayesian}.

Only changes in entropy are meaningful, and therefore entropy is defined with respect to a reference spectral function, the default model $d(\omega)$. This function eliminates the dependence on the choice of frequency grid or other fitting parameters. The entropy term is given by
\begin{equation}
S[A] = -\int d\omega\thinspace A(\omega)\ln\left[\dfrac{A(\omega)}{d(\omega)}\right].
\end{equation}

When using a default model, considering a range of $\alpha$ values becomes favorable, for example with: $\alpha\gg1$, we obtain the default model, which is an attempt to ensure a sensible solution that is ideally independent of the input data. For  $\alpha\ll1$, we again recover the least squares fit. While the default model provides a starting estimate  $A(\omega)$, the final result  is often insensitive to the choice of default model. 

This formalism can also be motivated in an equilibrium statistical mechanics formalism, where the free energy $F$ will be minimized. Since $F~=~U-TS$ there is interplay between internal energy $U$ and entropy $S$.
$Q$ can therefore be thought of as a quasi-free-energy where the parameter $\alpha$, much like $T$, contributes to the interplay between $U$ and $S$, where $U$ is represented by $\dfrac{1}{2}\chi^{2}$. 

\subsection{Algorithms for the Maximum Entropy Method}
The $Q$ functional of Eq.~\eqref{eq:Q} requires one to systematically determine the Lagrange multiplier, $\alpha$, and spectral function, $A$. Ref.~\citep{jarrell1996bayesian} lists two methods of determining a spectral function. The first, \emph{classic maxent} (in the terminology in Ref.~\citep{jarrell1996bayesian}), uses Bayesian inference to determine a spectral function that maximizes the function's posterior probability. The other method, \emph{Bryan's method} \cite{bryanMaxent}, instead takes all spectral functions found for a given range of $\alpha$ values and averages them by their posterior probability \citep{bryanMaxent}.
Since \emph{classic maxent} is a subset of \emph{Bryan's method}, this allows for both to be computed within the same run of a program. 

Bryan also provides an algorithm for reducing the search space of the kernel, by use of single value decomposition (SVD). For all of the kernels implemented in \ME, the eigenvalues of the kernel drastically drop off in magnitude providing a search space generally between 8-20 dimensions. Additonally, Bryan gives the formalism to minimize $Q$ using a Levenberg-Marquardt routine \citep{marquardt}. The code \MEsp implements Bryan's algorithm as described in Ref.~\cite{jarrell1996bayesian,bryanMaxent}.

Other algorithms to find solutions using the MEM exist, including those that interpret the entropy curve \citep{sandvik:1998} or finding a kink in a double-log plot of $\chi^2$ \citep{beach:2004} in order to optimize $\alpha$. The MEM has also been shown to be a special limit of the stochastic analytic continuation algorithm \cite{beach:2004}.  

\subsection{Error Estimates}
In addition to performing the inversion, $\ME$ is able to obtain systematic error bars that quantify variation in the \emph{classic maxent} solution. In Ref.~\citep{jarrell1996bayesian}, Jarrell and Gubernatis give an analysis of the curvature of the objective $Q$. They provide a second order approximation to the covariance matrix $\left\langle \delta A_{i}\delta A_{j}\right\rangle$ of the most probable spectral function. Using this covariance matrix, we have designed a routine to rotate \:$ \sqrt[]{A}$ into a diagonal basis, bootstrap within that basis, and rotate back to determine the error bar on the spectral function $A$. 

The fitting routine itself and the dependence on the default model can be interpreted as an alternate source of error. This should give some semblance to the systematic error in the fitting routine, however it underestimates the error, except for data that has a strong default model dependence where the dominant error becomes the choice of default model. To aid in understanding the default model dependence, \MEsp can run sequential calculations given choices of default model. 

\section{Implementation}
All \MEsp program options can be listed with a standard {\tt --help} option, and these are elaborated upon in the code documentation.  We note the most common user control parameters and pertinent details in this section.

\subsection{Input Basis}
There are three choices for input basis: imaginary time, frequency, and Legendre. Imaginary time consists of a real Green's function on $\tau\in[0,\beta]$. The frequency basis is a complex valued Green's function consisting of Masubara frequencies defined in the usual way. 
 
When the spectral function is symmetric about the Fermi level (generally taken to be $\omega=0$), the Green's function is known to have particle-hole symmetry. For fermionic Green's functions in the Matsubara basis, this corresponds to a vanishing real part, or vanishing imaginary part for bosonic Matsubara data. This allows us to only consider the real/imaginary part of the kernel, as shown in Table~\ref{tab:frequency-kernels}.

\MEsp supports, in addition to Green's functions, self-energies \cite{WangMillis09} and correlation functions. The self-energies and correlation functions do not generally conform to the same symmetries and normalization properties as Green's functions, so \MEsp  multiplies by the proper normalization so that the effective spectral function is entirely positive. Normalization can also be provided by the user, for cases where the total spectral weight is not unity. Once the effective spectral function has been calculated, \MEsp reverses the normalization in the output. 

This code also implements a fermionic Legendre representation of the Green's function. As formulated by Boehnke, et al. \citep{PhysRevB.84.075145}, a Green's function can be expanded in the basis of Legendre polynomials given by the transforms
\begin{equation}
G(\tau)=\sum_{\ell}\dfrac{\sqrt{2\ell+1}}{\beta}P_{\ell}\left[x(\tau)\right]G_{\ell},
\end{equation}

\begin{equation}
 G_{\ell}=\sqrt{2\ell+1}\int_{0}^{\beta}d\tau\thinspace P_{\ell}\left[x(\tau)\right]G(\tau),
\end{equation}
where $P_\ell$ is the $\ell$th Legendre polynomial and $x(\tau)=2\tau/\beta-1\in[-1,1]$. For particle-hole symmetric Legendre data, odd $\ell$ values will vanish. 

Representing a Green's function in an orthogonal polynomial basis, specifically the Legendre basis, has two major benefits over the time or frequency representation. Because of the density of information, the storage size for a Green's function is significantly reduced \citep{zgidLegendre}. This is advantageous when storing Green's function objects for large systems. The other benefit is that the kernel is generally at most on the order of about $40\times N_{\omega}$. For a small number of $\omega$ grid points $N_{\omega}$, $\ME$\! will perform considerably faster compared to other data representations.  This becomes important for general many-body problems where a Green's function has another index dependence, {\it e.g.} orbital number or momentum $\mathbf{k}$, in addition to $\tau$ or $i\omega_n$.  In this case, one must perform separate $\ME$ runs for each index point. 

\subsection{Default Models and Grids}
To generate the real frequency grid of the output spectral function, a user defined default model can be used (where the grid is taken from the defined points of the model), or a default model and grid can be generated with program options. The default models and grids  are defined explicitly in Tables \ref{tab:default_models} and \ref{tab:grids} in terms of their default and user required parameters, all of which are mutable via program options. The spectral function is then evaluated only on these grid points.

\renewcommand{\arraystretch}{2}
\begin{table*}[h]
\noindent

\begin{tabular}{|c|c|c|}
\hline 
\multirow{2}{*}{Name} & \multirow{2}{*}{Expression} & \cellcolor{lightgray}Required Variables\tabularnewline
\cline{3-3} 
 &  & Default Variables\tabularnewline
\hline 
\hline 
\multicolumn{3}{|c|}{Default Models $d(\omega)$}\tabularnewline
\hline 
Flat & $1/(\omega_{max}-\omega_{min})$ & --\tabularnewline
\hline 
\multirow{2}{*}{Gaussian} & \multirow{2}{*}{$\frac{1}{\sqrt{2\pi}\sigma}\exp\left[-\omega^{2}/(2\sigma^{2})\right]$} & \cellcolor{lightgray}$\sigma=\texttt{SIGMA}$\tabularnewline
\cline{3-3} 
 &  & --\tabularnewline
\hline 
\multirow{2}{*}{Double Gaussian} & \multirow{2}{*}{$\frac{1}{2}\left(\frac{1}{\sqrt{2\pi}\sigma}e^{-\frac{(\omega-\texttt{SHIFT})^{2}}{2\sigma^{2}}}+\frac{1}{\sqrt{2\pi}\sigma}e^{-\frac{(\omega+\texttt{SHIFT})^{2}}{2\sigma^{2}}}\right)$} & \cellcolor{lightgray}$\sigma=\texttt{SIGMA}$\tabularnewline
\cline{3-3} 
 &  & $\texttt{SHIFT}=0$\tabularnewline
\hline 
\multirow{2}{*}{Two Gaussians} & \multirow{2}{*}{$\frac{\texttt{NORM1}}{\sqrt{2\pi}\sigma1}e^{-\frac{(\omega-\texttt{SHIFT1})^{2}}{2\sigma1^{2}}}+\frac{(1-\texttt{NORM1})}{\sqrt{2\pi}\sigma2}e^{-\frac{(\omega-\texttt{SHIFT2})^{2}}{2\sigma2^{2}}}$} & \cellcolor{lightgray}$\texttt{SIGMA1,SIGMA2,SHIFT2}$\tabularnewline
\cline{3-3} 
 &  & $\texttt{NORM1}=0.5$, $\texttt{SHIFT1}=0$\tabularnewline
\hline 
\multirow{2}{*}{Shifted Gaussian} & \multirow{2}{*}{$\frac{1}{\sqrt{2\pi}\sigma}\exp\left[-(\omega-\texttt{SHIFT})^{2}/(2\sigma^{2})\right]$} & \cellcolor{lightgray}$\sigma=\texttt{SIGMA}$\tabularnewline
\cline{3-3} 
 &  & $\texttt{SHIFT}=0$\tabularnewline
\hline 
  \multirow{2}{*}{Lorentzian} & \multirow{2}{*}{$\left.1\middle/\left(\pi\gamma\left[1+\left(\frac{\omega}{\gamma}\right)^{2}\right]\right)\right.$} & \cellcolor{lightgray}$\gamma=\texttt{GAMMA}$\tabularnewline
\cline{3-3} 
 &  & ---\tabularnewline
\hline 
\multirow{2}{*}{LinearRiseExpDecay} & \multirow{2}{*}{$\lambda^{2}\omega\exp\left[-\lambda\omega\right]$} & \cellcolor{lightgray}$\lambda=\texttt{LAMBDA}$\tabularnewline
\cline{3-3} 
 &  & --\tabularnewline
\hline 
\multirow{2}{*}{QuadraticRiseExpDecay} & \multirow{2}{*}{$\dfrac{\lambda^{3}}{2}\omega^{2}\exp\left[-\lambda\omega\right]$} & \cellcolor{lightgray}$\lambda=\texttt{LAMBDA}$\tabularnewline
\cline{3-3} 
 &  & --\tabularnewline
\hline 
\end{tabular} \\

\protect\caption{Default model $d(\omega)$ names and expressions that are available in \protect\ME. \label{tab:default_models} }
\end{table*}
\begin{table*}[h]
\noindent

\begin{tabular}{|c|c|c|}
\hline 
\multirow{2}{*}{Name} & \multirow{2}{*}{Expression} & \cellcolor{lightgray}Required Variables\tabularnewline
\cline{3-3} 
 &  & Default Variables\tabularnewline
\hline 
\hline 
\multicolumn{3}{|c|}{Grids}\tabularnewline
\hline 
\multirow{2}{*}{Lorentzian} & \multirow{2}{*}{$t_{i}=\tan\left[\pi\dfrac{i}{\texttt{NFREQ}}(1-2\texttt{CUT})+\texttt{CUT-0.5}\right]$} & --\tabularnewline
\cline{3-3} 
 &  & $\texttt{CUT=0.01}$\tabularnewline
\hline 
\multirow{2}{*}{Half-Lorentzian} & \multirow{2}{*}{$t_{i}=\tan\left[\pi\dfrac{i+\texttt{NFREQ}}{2\texttt{NFREQ}+1}(1-2\texttt{CUT})+\texttt{CUT-0.5}\right]$} & --\tabularnewline
\cline{3-3} 
 &  & $\texttt{CUT=0.01}$\tabularnewline
\hline 
\multirow{2}{*}{Quadratic} & \multirow{2}{*}{$\Delta t=\dfrac{4(\texttt{SPREAD-1)}\cdot\left[(\dfrac{i}{\texttt{NFREQ}})^{2}-\dfrac{i}{\texttt{NFREQ}}\right]+\texttt{SPREAD}}{\texttt{NFREQ}/\left[3(\texttt{NFREQ}-1)\left(\texttt{NFREQ}\cdot(2+\texttt{SPREAD}\right)-4+\texttt{SPREAD}\right]}$} & --\tabularnewline
\cline{3-3} 
 &  & \multirow{1}{*}{$\texttt{SPREAD}=4$}\tabularnewline
\hline 
\multirow{2}{*}{Log} & \multirow{2}{*}{$t_{i}\sim0.5\pm\texttt{LOG\_MIN}\exp\left[i\cdot\log\left(0.5/\texttt{LOG\_MIN}\right)/(\texttt{NFREQ}/2-1)\right]$} & --\tabularnewline
\cline{3-3} 
 &  & $\texttt{LOG\_MIN}=0.0001$\tabularnewline
\hline 
Linear & $t_{i}=\dfrac{i}{\texttt{NFREQ}}$ & --\tabularnewline
\hline 
\end{tabular}\\

\protect\caption{Grid names and expressions that are available in \protect\ME. A real frequency grid is generated with \protect$t_i\in[0,1]$ and then mapped to the range \protect$\omega_i\in[\omega_{min},\omega_{max}]$. The bounds for the frequency grid are controlled through the parameters  \protect\texttt{OMEGA\_MAX} and \protect\texttt{OMEGA\_MIN}. \label{tab:grids} }
\end{table*}

\subsection{Kernels}\label{sec:kernels}
\MEsp implements several common kernel choices, for fermionic, bosonic, and anomalous Green's functions with and without particle-hole symmetry. These are shown in Tables \ref{tab:frequency-kernels} and \ref{tab:time-kernels}. After choosing the input basis and default model (with $\omega$ grid), the kernel is automatically set up. There is an additional problem with the bosonic kernel, where the kernel is not only singular for $\omega=0$ and $n=0$ but the spectral function is negative below the Fermi level. To overcome this, \MEsp uses the kernel 
\begin{equation}
K_n(\omega)=\dfrac{\omega}{i\omega_n+\omega}
\end{equation}
which relates to the effective function $B(\omega)=\chi^{\prime\prime}(\omega)/\omega$; after the calculation of $B(\omega)$, \MEsp also produces $\chi^{\prime\prime}(\omega)=\omega B(\omega)$ \citep{jarrell1996bayesian}.  

For a Green's function represented by $G_\ell$ in the Legendre basis, the representation of Eq.~\eqref{eq:G=KA} can be written with a kernel
\begin{equation}
K_\ell(\omega)\equiv-\sqrt{2\ell+1}\int_{-1}^{1}dx\thinspace\dfrac{e^{-(1+x)\beta\omega/2}}{1+ e^{-\beta\omega}}P_{\ell}(x).
\end{equation}
\MEsp uses GSL to integrate the Legendre kernel \citep{GSL}.

\begin{table}[h]
\noindent
\begin{tabular}{|c|c|c|}
\hline 
Dataspace & Kernel Name & Kernel\tabularnewline
\hline 
\hline 
Frequency & Fermionic  & $\dfrac{1}{i\omega_{n}-\omega}$\tabularnewline
\cline{2-3} 
\multirow{2}{*}{Without PH} & Bosonic & $\dfrac{\omega}{i\omega_{n}+\omega}$\tabularnewline
\cline{2-3} 
 & Anomalous & $\dfrac{-\omega}{i\omega_{n}-\omega}$\tabularnewline
\hline 
Frequency & Fermionic & $-\dfrac{\omega_{n}}{\omega_{n}^{2}+\omega^{2}}$\tabularnewline
\cline{2-3} 
With PH & Bosonic & $\dfrac{\omega^{2}}{\omega_{n}^{2}+\omega^{2}}$\tabularnewline
\cline{2-3} 
 & Anomalous & $\dfrac{\omega^{2}}{\omega_{n}^{2}+\omega^{2}}$\tabularnewline
\hline 
\end{tabular}

\caption{Kernels in Matsubara frequency\label{tab:frequency-kernels}}
\end{table}

\begin{table}[h]
\noindent
\begin{tabular}{|c|c|}
\hline 
Kernel Name & Kernel\tabularnewline
\hline 
\hline 
Fermionic & $-\dfrac{e^{-\tau\omega}}{1+e^{-\omega\beta}}$\tabularnewline
\hline 
Bosonic & $\frac{1}{2}\omega\dfrac{\left[e^{-\omega\tau}+e^{-\omega(\beta-\tau)}\right]}{1-e^{-\omega\beta}}$\tabularnewline
\hline 
TZero & $-e^{-\omega\tau}$\tabularnewline
\hline 
\end{tabular}

\caption{Kernels in imaginary time\label{tab:time-kernels}}
\end{table}

\section{Examples}
We provide four detailed examples for the different data representations available to \ME. 

In order to provide physically motivated examples, we include example data generated for a relevant model of correlated systems, the Hubbard model on a square lattice in 2-dimensions, which has been the topic of rigorous numerical benchmarking \cite{benchmarks}.

The Hamiltonian of the interacting Hubbard model can be written as
\begin{equation}
H=-\sum_{<ij>\sigma}t\left(c_{i\sigma}^{\dagger}c_{j\sigma}+c_{j\sigma}^{\dagger}c_{i\sigma}\right) + U\sum_i n_{i\uparrow}n_{i\downarrow}
\label{eq:interacting-hubbard}
\end{equation}
where $c,c^\dagger$ are the annihilation/creation operators and the sum is over nearest neighbor sites $i,j$ with spin $\sigma$ and 	hopping amplitude $t$. Solutions to the Hubbard model using dynamical mean field theory (DMFT) will undergo a transition to a Mott insulator upon increasing the ratio of $U/t$ at half-filling \cite{georges:1996,georges:1993,jarrell:1992}.

\subsection{Fermionic Green's Function}
Here we present a walk-through of the non-interacting Hubbard Model of Eq.~\eqref{eq:interacting-hubbard} at $U/t=0,\: \beta t=8$. The problem can be solved analytically for $U=0$.  One can then generate a local Green's function in Matsubara frequencies. All example data unless otherwise noted was calculated at half-filling, and thus only the imaginary part is needed as input.

For this example, the frequency space input data is in the file ``G\_im,'' which has columns of the form ``$i\omega_n \, \textvisiblespace \, \text{Im}[G_n]\, \textvisiblespace \, \sigma_n$'', where the set of $\sigma_n$ are simulated errors chosen to be a fixed small value to avoid an over constrained fitting routine. The parameter input file provided to \MEsp is:

\begin{lstlisting}[language=Python,basicstyle={\small\ttfamily},frame=tb,title={Param File in.param},escapechar={!}]
BETA=8			#inverse temperature
NDAT=1024		#num of data points
NFREQ=500               #num of output frequencies
DATASPACE=frequency	#!$G(i\omega)$!
KERNEL=fermionic         #fermionic|bosonic values	
PARTICLE_HOLE_SYMMETRY=1 #0|1 
DATA="G_im"		 #location of data file
\end{lstlisting}
Running \texttt{maxent --help} will list default parameters used for this simulation.

\subsubsection{Output Guide}
\MEsp produces useful pieces of output during its calculations which are discussed in the example documents found in the folder ``examples.'' The SVD of this data set reduces the kernel to a singular space of 8 vectors, which produces a most probable spectrum and Bayesian averaged spectrum with 60 $\alpha$ values in the range $[0.01,20]$. The normalization of the spectral function varies from unity by no more than $3\times 10^{-4}$ and when back-continued, the two spectra have an error no larger than $4\times10^{-4}$ --- both of which are a sign of a successful calculation. 

\begin{figure}[h]
\noindent \begin{centering}
\protect\includegraphics[viewport=0bp 0bp 750bp 540bp,clip,scale=0.3]{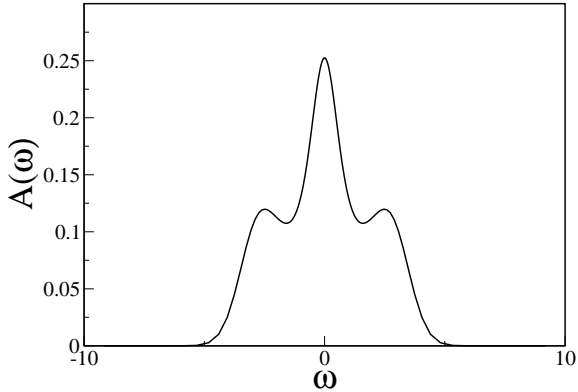} 
\protect\caption{Spectral function from $\protect\ME$ (Bryan's method) using
data at $U/t=0$, $\beta/t=8$\label{fig:Spectral-function}}
\end{centering}
\end{figure}

If the parameter \texttt{TEXT\_OUTPUT} is set to true, \MEsp will produce eight files described in Table~\ref{fig:ME-filenames}. The main output is the spectral function obtained using Bryan's method, which is written to \emph{in.out.avspec.dat} (given an input file named \emph{in.param} - see Table~\ref{fig:ME-filenames}) and is shown in Figure \ref{fig:Spectral-function}. 

\begin{table*}[h]
\noindent
\renewcommand{\arraystretch}{1}
\begin{tabular}{|ll|}
\hline 
\code{name.out.\uline{avspec}.dat} & Spectral function using Bayesian Averaging - \textbf{Bryan's method}\tabularnewline
\code{name.out.\uline{avspec}\_back.dat} & The back-continued \code{avspec} spectrum \tabularnewline
\code{name.out.\uline{chi2}.dat } & Estimated $\chi^{2}$ for each $\alpha$ value solution\tabularnewline
\code{name.out.\uline{chispec}.dat} & Spectral function satisfying the best $\chi^{2}$ - historic maxent\tabularnewline
\code{name.out.\uline{chispec}\_back.dat} & The back-continued \code{chispec} spectrum\tabularnewline
\code{name.out.\uline{fits}.dat} & Fits of each $\alpha$ value, see comments in file\tabularnewline
\code{name.out.\uline{maxspec}.dat}  & Spectral function with the highest probability - \textbf{classic maxent}\tabularnewline
\code{name.out.\uline{maxspec}\_back.dat} & The back-continued \code{maxspec} spectrum \tabularnewline
\code{name.out.\uline{out.h5}} & All output data in the hdf5 format\tabularnewline
\code{name.out.\uline{prob}.dat} & The posterior probability of each $\alpha$ value\tabularnewline
\code{name.out.\uline{spex}.dat} & All spectral functions produced; one for each $\alpha$\tabularnewline
\hline 
\end{tabular}

\protect\caption{Output files from \protect\MEsp given an input file \code{name.param}. The prefix \code{name} is replaced with the basename of the input file, or specified with the parameter \code{BASENAME}. \label{fig:ME-filenames}}
\end{table*}

\subsection{Fermionic Self-Energy}
 Using DMFT to solve a single site impurity problem on a square lattice, one can generate a local Green's function and self-energy for a metallic phase ($U/t=1$) and an insulating phase ($U/t=10$). Further explanation and datasets are available in the repository. 

\begin{figure}[h]
\noindent \begin{centering}
\subfloat[]{\protect\includegraphics[scale=0.30]{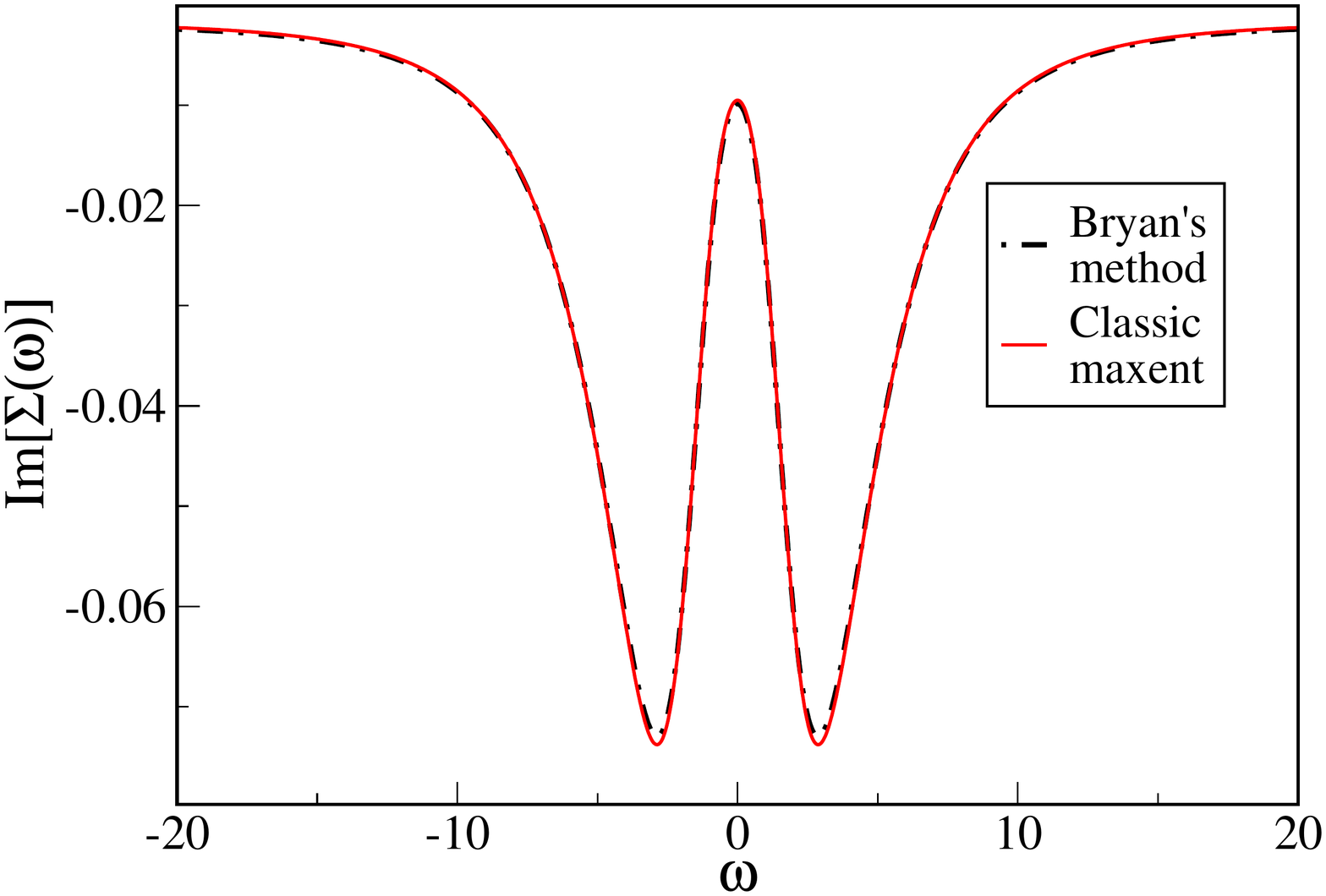}
}

\subfloat[]{\protect\includegraphics[scale=0.30]{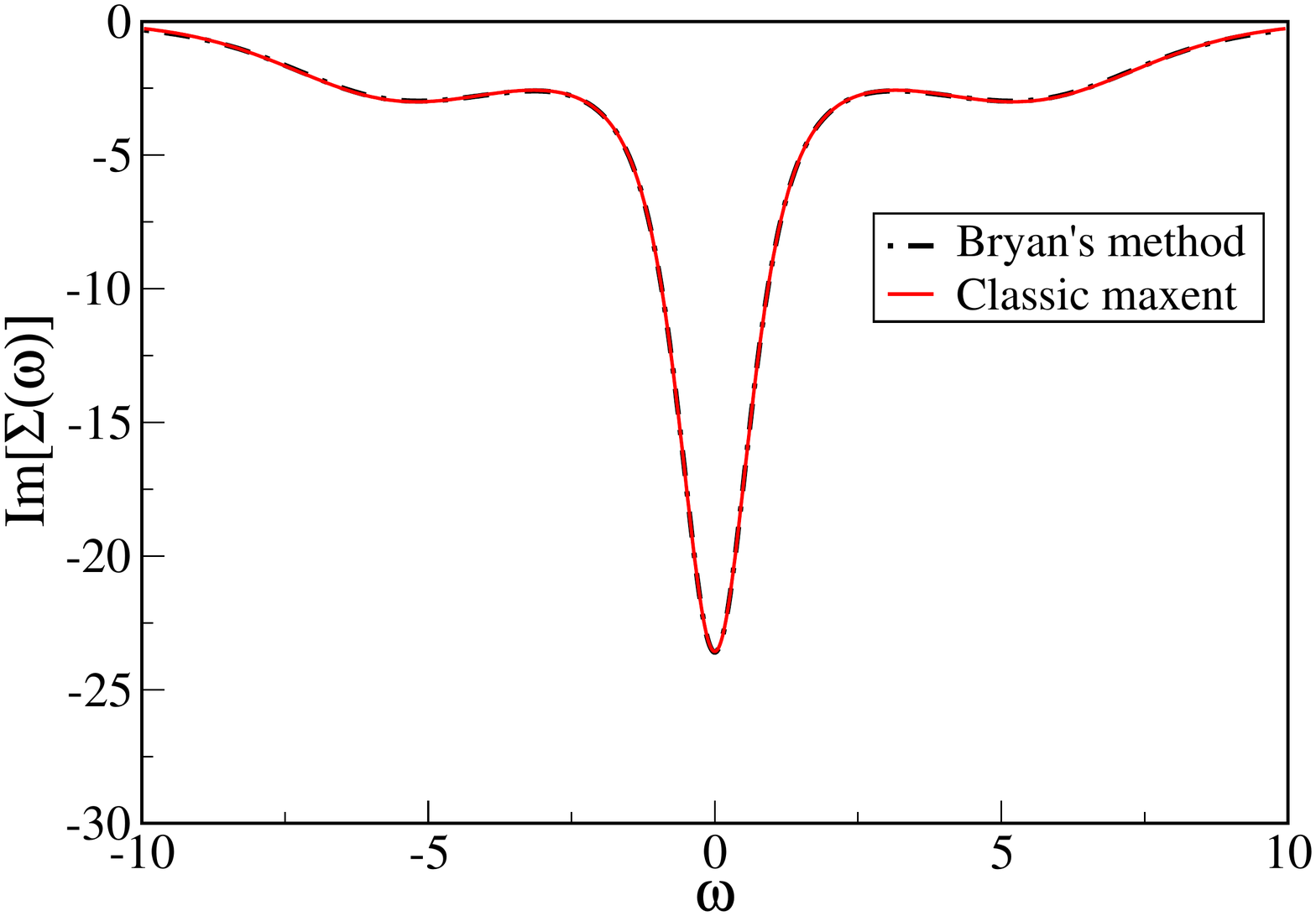}
}

\protect\caption{\protect\MEsp output of the imaginary part of the self-energy at $\beta t=2$ for: (a) $U/t=1$. (b) $U/t=10$.  Black curves are from Bryan's method and red curves from historic Maxent \label{fig:Self-Energy-ME}}
\end{centering}
\end{figure}

To activate the self-energy continuation in \ME, the parameter \texttt{SELF} needs to be set to true. This produces files \emph{in.out.avspec\_self.dat} and \emph{in.out.maxspec\_self.dat} which are the properly normalized self-energies for those respective functions. The self-energy output of \MEsp is shown in Figure \ref{fig:Self-Energy-ME}a-b.

\subsection{Legendre Representation}
\begin{figure}[h]
\noindent \begin{centering}

\subfloat[]{\protect\includegraphics[scale=0.30]{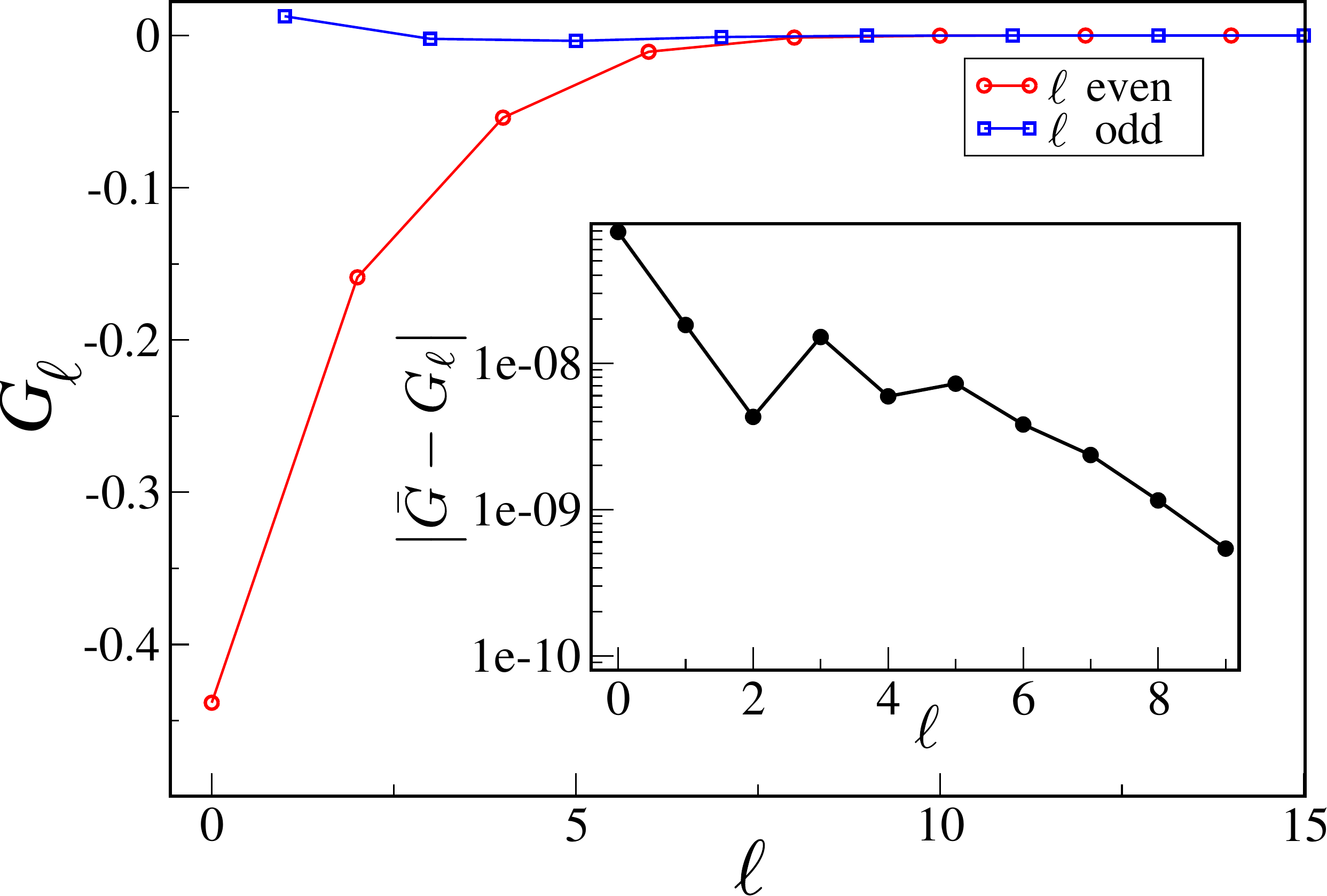}
} 

\subfloat[]{\protect\includegraphics[scale=0.30]{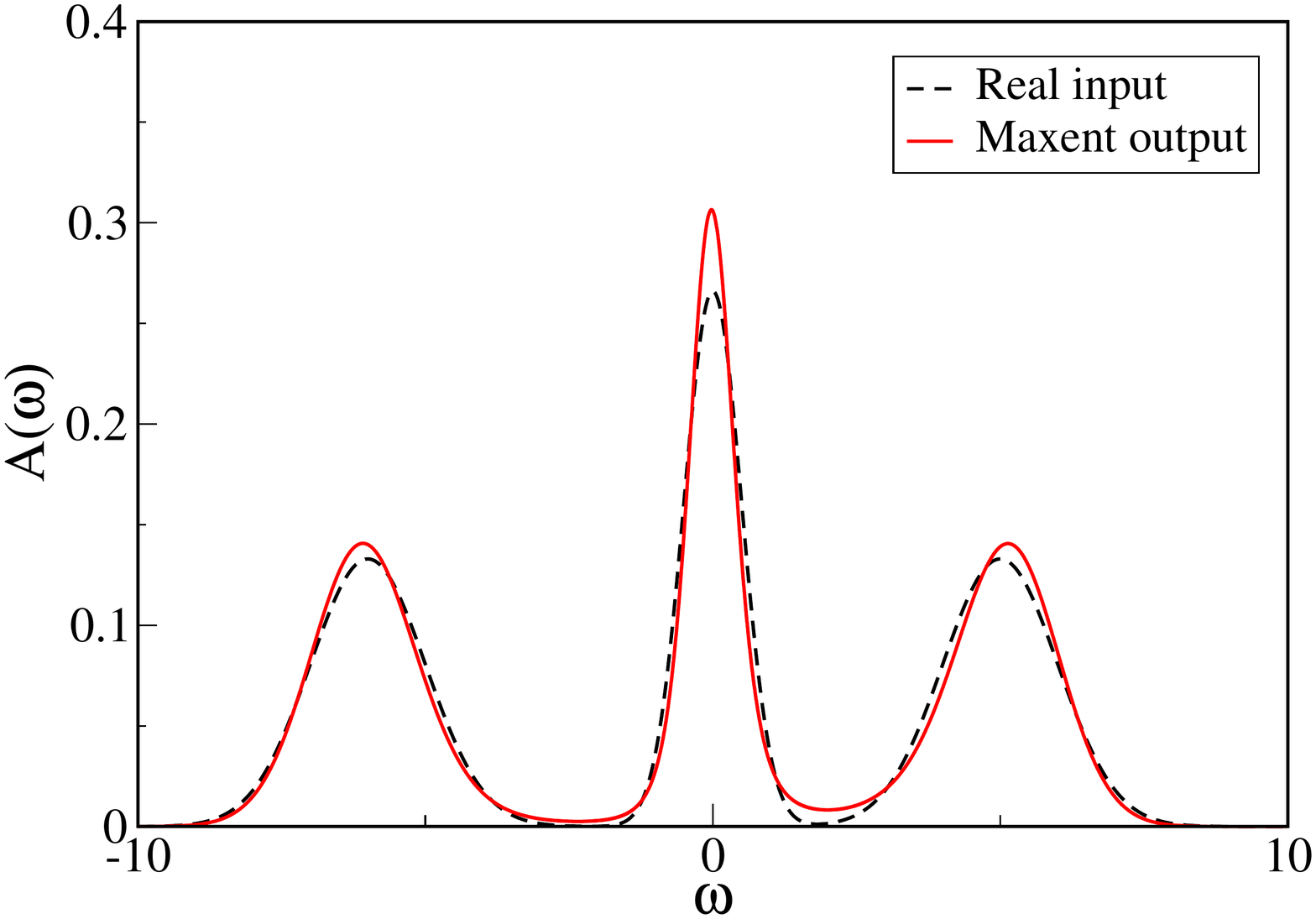}
}

\protect\caption{Legendre input and output at $\beta t=2$. (a) Legendre basis representation of input data. Inset: Error from of the spectral function on the Legendre axis. (b) Output from \protect\ME \label{fig:Legendre-plots}}
\end{centering}
\end{figure}

In order to provide an example for the Legendre kernel in \ME, a spectral function consisting of three Gaussian functions was calculated, back-continued into a fermionic imaginary time Green's function, and finally transformed into the Legendre basis.  The Legendre representation of the data is shown in Figure \ref{fig:Legendre-plots}a. To use a Green's function in the Legendre basis, change the parameter \texttt{DATASPACE} to \-``Legendre'' and keep \texttt{KERNEL} set to ``fermionic". Using only the first 10 points, \MEsp was able to produce an output shown in Figure \ref{fig:Legendre-plots}b, and when back-continued has absolute error shown in the inset of Figure  \ref{fig:Legendre-plots}a.

\subsection{Bosonic Green's Function}\label{sec:bosonic}
The Hubbard model in Eq.~\eqref{eq:interacting-hubbard} can also be solved for 2 particle correlation functions. 
We apply the dynamical cluster approximation on an 8-site cluster to obtain the magnetic susceptibility, $\chi (i\omega_n)$, as a function of bosonic Matsubara frequencies for the two-dimensional Hubbard model at $U/t=6,\: n=0.9,\:\beta t=2$ at a scattering momentum of $Q=(\pi,\pi)$. 

\begin{figure}[h]
\noindent \begin{centering}
\subfloat[]{\protect\includegraphics[scale=0.30]{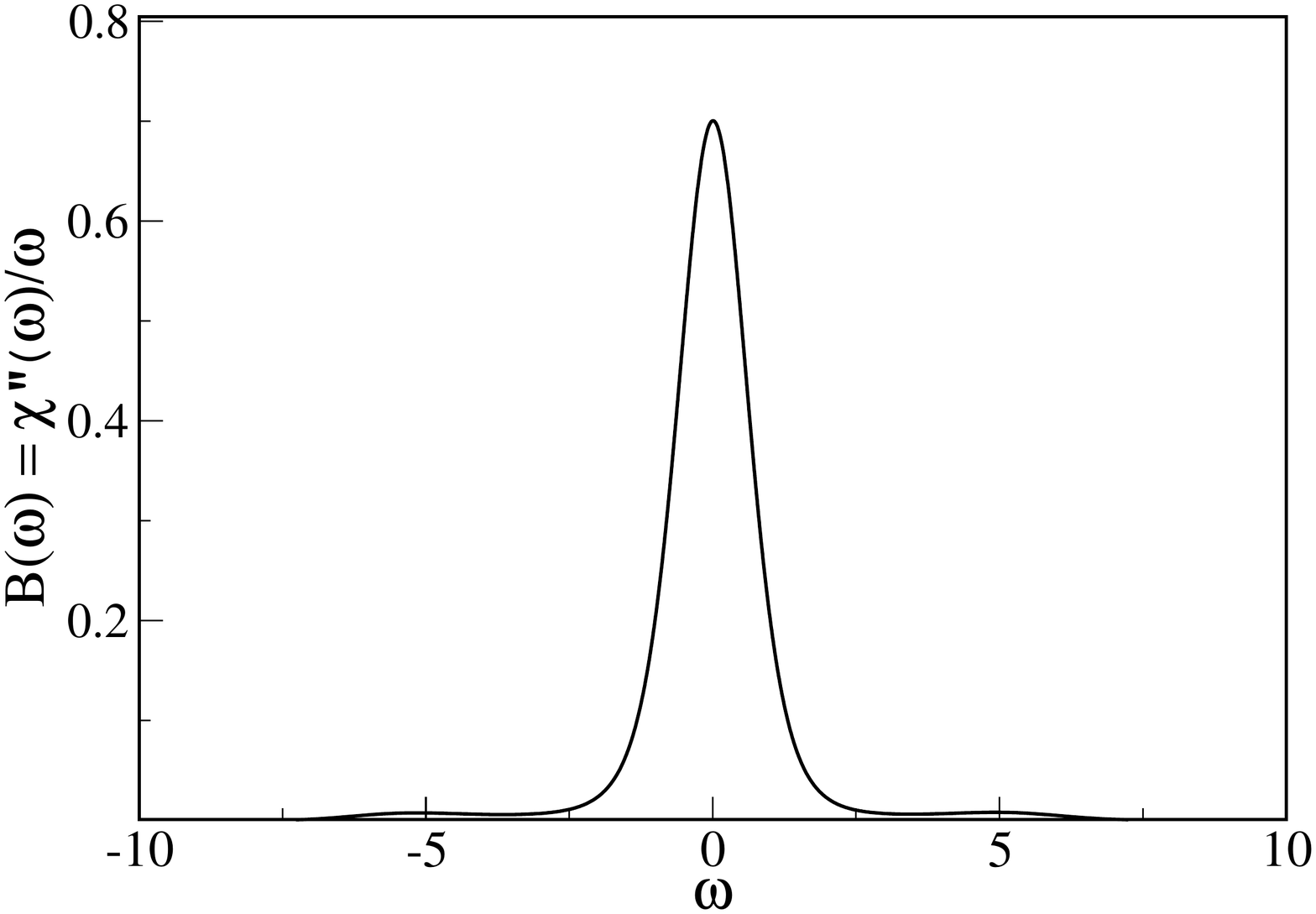}
}

\subfloat[]{\protect\includegraphics[scale=0.30]{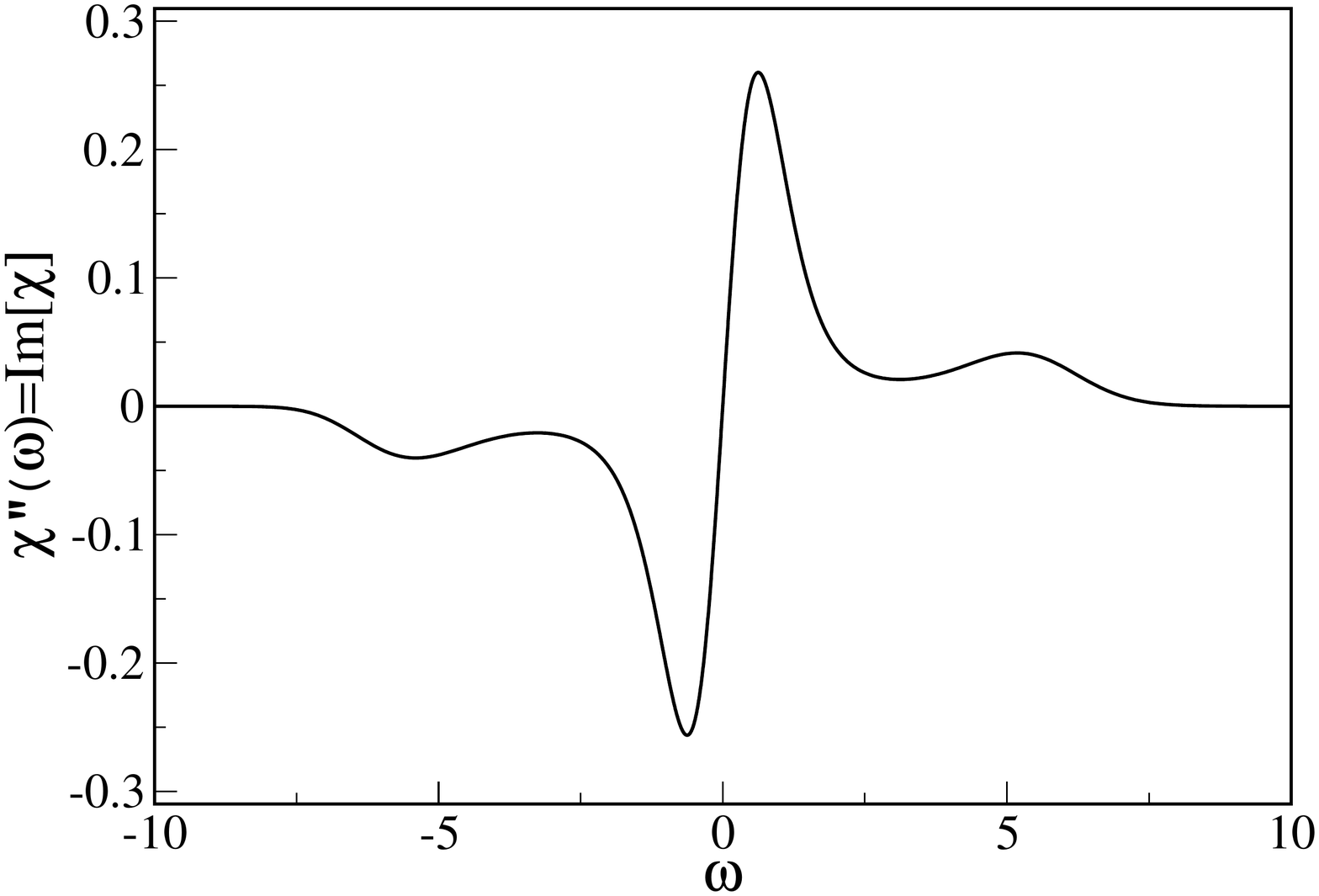}
}

\protect\caption{\protect\MEsp output for the magnetic susceptibility using bosonic Kernal. (a) Normalized function \protect$B(\omega)=\chi^{\prime\prime}(\omega)/\omega$. (b) The imaginary part of the susceptibility \protect$\chi^{\prime\prime}(\omega)$. \label{fig:Bosonic-Plot}}
\end{centering}
\end{figure}

To perform \MEsp on bosonic data, the parameter \texttt{KERNEL} must be set to \-``bosonic". \MEsp will produce extra files \emph{in.out.avspec\_bose.dat} and \emph{in.out.maxspec\_bose.dat} which are the properly normalized $\chi^{\prime\prime}(\omega)$, where in the usual spectral output, \emph{in.out.avspec.dat} for instance, will be $B(\omega)$. The results of both $\chi^{\prime\prime}(\omega)$ and $B(\omega)$ from \MEsp are shown in Figure \ref{fig:Bosonic-Plot}a-b. 

\section{Acknowledgments}
This project was supported by the Simons Collaboration on the Many-Electron Problem. We gratefully acknowledge assistance from the ALPSCore community and Alex Gaenko. 

\bibliographystyle{elsarticle-num}

\bibliography{maxent}

\end{document}